\documentstyle[prl,twocolumn,aps]{revtex} 
\begin{document} 
\draft
\title{Quantum ergodicity and localization 
in conservative systems:\\ 
the Wigner Band Random Matrix model} 
 
\author{G.Casati$^1$, B.V.Chirikov$^{1,2}$, I.Guarneri$^1$ and F.M.Izrailev$^ 
{1,2}$} 
\address{
$^1$ Universit\'a di Milano, Sede di Como, 
Via Lucini 3, 22100 Como, Italy\\
$^2$ Budker Institute of Nuclear Physics, 630090 Novosibirsk, Russia
}
 
\twocolumn[
\maketitle
\widetext
\vspace*{-1.0truecm}
\begin{abstract} 
\begin{center}
\parbox{14cm}{
First theoretical and numerical results on the global structure of the energy 
shell, the Green function spectra and the eigenfunctions, both localized and  
ergodic, in a generic conservative quantum system are presented. 
In case of quantum localization the eigenfunctions are shown to be typically 
narrow and solid, with centers randomly scattered within the semicircle energy shell while 
the Green function spectral density (local spectral density of states)
is extended over the whole shell, 
but sparse.} 
\end{center}
\end{abstract}
\pacs{
\hspace{1.9cm} 
PACS number 05.45.+b}
]

\narrowtext

One of the main results in the study of the so-called quantum chaos has been  
 the  
discovery of quantum dynamical localization as a mesoscopic quasi-classical 
phenomenon \cite{bible}. 
This phenomenon has been widely studied and confirmed by many researchers for  
dynamical models described by maps. Contrary to a common belief, maps  
describe not only time-dependent systems, but also conservative ones (in the  
form of Poincare' maps). On the other hand, to our knowledge, there are no  
direct studies  
of quantum dynamical localization in bounded conservative models; 
moreover, the appearance of dynamical localization in such systems 
due to quantum effects is challenged by some 
researchers. 
The existence of localization in conservative systems would restrict quantum 
distributions to smaller regions of phase space than classically allowed,  
and would therefore introduce significant deviations from ergodicity. 
 
We have addressed this problem on the Wigner Band Random Matrix (WBRM) model, 
which was introduced by Wigner  40 years ago \cite{W}  
for the description of complex, conservative quantum systems like atomic nuclei. 
Due to severe mathematical difficulties, the random matrix theory (RMT) 
 immediately turned to the much simpler case of statistically homogeneous 
(full) matrices, for which impressive theoretical results have been achieved 
(see, e.g., Refs.\cite{BM}). However, full matrices  describe local 
chaotic structures only, and this limitation is often inacceptable, for  
instance in the case of atoms \cite{Ch,FG}. 
 
Generally speaking, RMT is a statistical theory of systems with discrete energy 
(and frequency) spectrum. Since the latter is a typical property  
of quantum 
dynamical chaos \cite{CaCh}, RMT provides a statistical description of 
 quantum 
chaos and, what is very important, one which does not involve  
any coupling to a  
thermal 
bath, which is a standard element in most statistical theories. 
Moreover, a {\it single} matrix from a given statistical ensemble 
represents the typical (generic) {\it dynamical} system of a given class, 
characterized by a few  matrix parameters. This makes an important bridge 
between dynamical and statistical description of quantum chaos. 
 
To the extent that Band Random Matrices can be taken as models for   
generic few--freedoms conservative systems which 
are classically strongly chaotic (in particular ergodic) on a compact energy 
surface, the results presented in this Letter provide the first  
characterization of the properties of quantum chaos in momentum space for  
quantum systems of this class.

We consider real Hamiltonian matrices of a rather general type (more specific  
random matrix models have been recently proposed in \cite{Flam}) 
\begin{equation} 
 H_{mn}\,=\epsilon_n\,\delta_{mn}\,+\,v_{mn}\qquad(m,n=1,..,N) 
\label{eqWm} 
\end{equation} 
where off--diagonal matrix elements $v_{mn}=v_{nm}$ are statistically 
independent, Gaussian random variables, with $<v_{mn}>=0$ and $<v_{mn}^2>=v^2$, if  
$|m-n|\leq b$, and are zero otherwise.  
In a classical picture, WBRMs like (\ref{eqWm}) would correspond to  classical  
Hamiltonians of the form: 
\begin{equation} 
 H\,=\,H_0\,+\,V 
\label{eqclH} 
\end{equation} 
where the perturbation $V$ is usually assumed to be sufficiently small, 
while the unperturbed Hamiltonian $H_0$ is completely integrable. 
In the quantum model the matrix (\ref{eqWm}) is given in the basis of  
the unperturbed eigenstates $\phi_n$ of ${\hat H}_0$. Correspondingly, the  
fluctuations of 
unperturbed energy levels $\epsilon_n$ are taken as Poissonian. Although in  completely  
integrable quantum system there is a quantum number for each freedom, we  
suppose that the  unperturbed states are ordered according to increasing  
energy, and we thereby label them by a single number $n$.  
 The most important  
characteristic of WBRM is the average level density $\rho$: 
\begin{equation} 
 \rho^{-1}\,=\,\left<\epsilon_n-\epsilon_{n-1}\right> 
\label{eqdl} 
\end{equation} 
Here and below, the averaging  is understood either over disorder (that is, over 
many random matrices) or within a single, sufficiently large, matrix. Both 
ways are equivalent owing to assumed independence of matrix elements.

In the classical case, the unperturbed energy $E_0$ is not constant along a classical chaotic  
trajectory of the full Hamiltonian with a given  
total energy $H=E$. Instead, it sweeps a range of values, or  
"energy shell",  
$ \Delta E_0 =\Delta V$, and is distributed inside this shell according  
to a measure  $W_{E}(E_0)$. The form of  $W_E(E_0)$  
 depends on the form of the perturbation $V$; we will call this measure 
"ergodic"  
because it is determined by the ergodic (microcanonical) measure on the  
given energy  
surface $H=E$.   
The quantum analog of this measure characterizes the distribution 
of the "ergodic" eigenfunction (EF) in the unperturbed basis.  
 
Conversely,  
if we keep the  
unperturbed energy $E_0$ fixed,  the bundle of trajectories of the  
total Hamiltonian $H$, which reach the surface $H_0=E_0$, has a distribution  
in the total energy $E$ which is described by a measure $w_{E_0}(E)$.  
In the quantum case,  
this measure  corresponds to the energy spectrum of the 
Green function (GFS) at energy $E_0$, and has received different  
names, such as "strength function", "local spectral density of states",  
"spectral measure" of the  
unperturbed eigenstate at energy $E_0$.  
 
An expression for the latter measure has been given by Wigner 
\cite{W}. For a typical perturbation, 
represented by a WBRM, the average measure $w(E)=\langle w_{E_0}(E)\rangle   
$ depends on the Wigner parameter  ,  
\begin{equation} 
 q\,=\,\frac{(\rho \,v)^2}{b} 
\label{eqq} 
\end{equation} 
and has the following limiting forms 
 \cite{W} (see also Refs.\cite{CaChCo,YaF,Fe}) 
\begin{equation} 
 w(E)\,=\,\left\{  
\begin{array}{ccc} 
 \frac{2}{\pi E_{sc}^2}\,\sqrt{E_{sc}^2\,-\,E^2}, &  
|E|\,\le\,E_{sc}, & q\,\gg\,1\\ 
 & & \\
 \frac{\Gamma /2\pi}{E^2\,+\,\Gamma^2/4}\cdot 
 \frac{\pi}{2\cdot\arctan{(1/\pi q)}}, & |E|\,\le\,E_{BW}, & q\,\ll\,1 
\end{array} \right. 
\label{eqBW} 
\end{equation} 
Outside the specified energy intervals, both distributions have exponentially 
small tails. Formulae (\ref{eqBW}) are valid 
provided $\rho v > 1$, which is the condition for strong coupling of neighboring  
unperturbed states by the perturbation. In the opposite  
case $\rho v<1$ the  
effect of the perturbation is small, and we have the so-called perturbative  
localization. 

In the limit $q>>1$ we have the semicircle (SC) law and the width of  
the energy shell 
$\Delta E = 2E_{sc}=4v\sqrt{2\,b}=4\sqrt{2q}E_b\gg E_b$ where $E_b=b/\rho$  
is the half width (in energy) of the band. In the other limit, $q<<1$,     
we have the Breit - Wigner (BW) distribution, of width  
$\Delta E = 2E_{BW}=2E_b$, with the main part inside a 
width $\Gamma = 2\pi\rho v^2 = 
2\pi qE_b\ll E_b$.  
In all these expressions $E$ is measured with respect to the center of 
the distribution. Since $q<<1$ requires $\rho v<\sqrt{b}$,  
in the BW regime  the perturbation is not strong enough to couple all states  
within one bandwidth. This means that the BW regime corresponds in fact 
to a sort  
of partial perturbative localization.  
 
The numerical results presented below are contained in 
the EF matrix  
$C_{mn}$, which connects exact eigenfunctions $\psi_m$, obtained 
by diagonalization of the Hamiltonian matrix (\ref{eqWm}), 
to unperturbed basis 
states  
$\phi_n$, 
\begin{equation} 
 \psi_m\,=\,\sum_n\,C_{mn}\cdot\phi_n  
\label{eqefm} 
\end{equation} 
In what follows the eigenvalues $E_m$ are ordered, so 
that $E_m\approx m/\rho$. 
 
From the matrix $C_{mn}$ we have found both the statistical distribution  
 $W_m(n)=C^2_{mn}$ of the eigenstates $\psi_m$ on the unperturbed ones 
$\phi_n$, and the  
distribution $w_n(m)$ of the unperturbed  
eigenstates on the exact ones; the meaning of these distributions is similar to  
that of the classical $W$ and $w$ discussed above. 
We have then analyzed both distributions, and have compared their  
structures to each other and to the SC distribution, paying  
special attention to localization.  
By localization we shall here mean a situation, in which eigenfunctions  
are localized on a scale which is significantly smaller than the  
maximum one consistent with energy conservation.  
Indeed,  
the size of the region which is populated by an eigenfunction (termed 
localization length in the following) is bounded from above by the  
{\it ergodic localization length} 
$d^{(e)}=c\rho\Delta E$, 
which measures the maximum number of basis states coupled by the perturbation. 
This length characterizes the full width of the energy shell $\Delta E$. The factor $c$  
depends on the definition of localization width (see eqn.(\ref{eqdhp}) below). 
In other words, in a conservative quantum system there is always  
localization in  
energy, due  
to the existence of a  
finite $\Delta E$ \cite{CCGI}. This fact, which is sometimes a source of   
 confusion  
,  is  just  a trivial consequence 
of energy conservation. 
Here we are interested in   
localization {\it inside} the shell \cite{CCGI}, which can be caused by  
 quantum effects. In this connection, the matrix size $N$ is  
an irrelevant parameter, provided $N\gg d^{(e)}$ 
is large enough to avoid boundary effects.  The quantum model (\ref{eqWm}) is 
thus defined by the 3 physical parameters 
$\rho ,\ v,$ and $b$. 
 
The localization length $d_m$ of a distribution $w_m(n)$ 
can be defined in several ways. We  
have used 
the so--called inverse 
participation ratio (see, e.g., Ref.\cite{CaCh}): 
\begin{equation} 
 d_{m}^{-1}\,=\frac13\sum_n\,w^2_m(n) 
\label{eqdhp} 
\end{equation} 
and similarly for $W_n(m)$. The numerical factor $1/3$ accounts for 
fluctuations in individual distributions. These distributions are assumed 
to be Gaussian and independent\cite{FG}.
 
 In order to suppress  
large fluctuations in individual distributions of both types $W_m(n)$ and $w_n(m)$
, we have taken  
averages over $300$ of them, chosen around the center of the spectrum. 
 Since different distributions cover different regions of the $n$  
(respectively, $m$) space, prior to averaging they have to be shifted into a common  
region. This we have done in two different ways, namely, either by counting   
the site label $n$ in $W_m(n)$ starting from the center of the energy shell  
i.e., from the  reference site $m$ (and vice- 
versa in the case of $w_n(m)$)(circles in Figs.1,2), or from the 
center $n_c(m)$ of $W_m(n)$,  
defined by  
\begin{equation} 
 n_c(m)\,=\,\sum_n\,W_m(n)\cdot n 
\end{equation} 
The two types of average will be denoted by $\langle W(n)\rangle, 
\overline {W}(n)$ respectively. In particular, $\overline{ W}(n)$  
yields the average shape of an eigenstate (full line in  
Figs.1,2). 

 First we shall discuss  the distributions $W_m(n)$. 
In \cite{CCGI} it was shown that  the average  
 localization length $d\equiv\langle (\sum_n W^2_m(n))^{-1}\rangle$
 obeys a scaling law of the form    
 \begin{equation} 
 \beta_d\,=\,\frac{d}{d^{(e)}}\,\approx\,1\,-\,{\rm e}^{-\,\lambda}\,<\,1 
\label{eqdde} 
\end{equation} 
where 
\begin{equation} 
\lambda\,=\,\frac{ab^2}{d^{(e)}}\,=\,\frac{ab^{3/2}}{4\sqrt{2}c\rho v} 
\label{erpar} 
\end{equation} 
Here $a\approx 1.2$; factor $c$ can be directly calculated from the limiting 
expression 
(\ref{eqBW}) for $w$, which gives $c\approx 0.92$.

The empirical relation (\ref{eqdde}) has been found \cite{CCGI} 
 to hold in the whole interval $\lambda\leq 2.5$ and was confirmed 
in the present studies up to $\lambda\approx 7$\cite{FLW}. 
 
The parameter $\lambda$ has been shown \cite{CCGI} 
to play the role of an {\it ergodicity parameter} because, 
when it is large, the localization length approaches its maximal value 
$d^{(e)}$, which means that the eigenfunctions become ergodic, i.e., delocalized 
over the whole energy shell.
Notice that in the BW region the ergodic localization 
length $ d^{(e)}=\pi\rho\Gamma= 
2\pi^2 b q$, and 
$\lambda\approx ab/2q\pi^2\gg 1$ \cite{CaChCo} since 
$q\ll 1$ (and $b\gg 1$ in quasiclassical situations)
 . Hence, localization is only possible 
in the parameter range in which the local density of states follows the 
SC law. This domain is the main object of the present studies. 
 
In the case $\lambda>>1$ (Fig.1a) we have found that  
the averaged distributions $\langle W(n)\rangle,\overline{W}(n)$ are 
 fairly close to the SC law: a remarkable result, because 
 that law was theoretically predicted for the {\it other} distribution, namely,  
for the GFS spectrum $\langle w(m)\rangle$. 
We presume that the deviations from  
the SC law which are observed in the 
distribution $\overline {W}(n)$ are due to  
the not very large value of the ergodicity parameter ($\lambda=3.7$).  
 The numerical values  
of the localization parameter (\ref{eqdde}) are $\beta=0.94$ and $\beta=1.08$ for the two  
types of average, respectively,   
in a reasonable agreement with 
the average $\beta_d=0.97$ computed from (\ref{eqdde}) for $\lambda =3.6$.  
 For finite $q$ the average distributions of both types are 
 bordered by two 
symmetric steep tails, which apparently fall down even faster than the 
simple exponential. The structure of these tails will be discussed in detail  
elsewhere. 
 
The structure of EFs is completely different in the case $\lambda<<1$. 
(Fig.1b). Whereas individual eigenstates exhibit  
large fluctuations, the main part of the average distribution ( 
with respect to the center $n_c$) $\overline{W}(n)$  shows  
a clear evidence for exponential localization, with localization length in  
agreement with the empirical formula (\ref{eqdde}). 
The width of the main part is small 
($ \beta=0.24$), which is again close to  
average $\beta_d=0.21$ for $\lambda =0.24$. 
We have found that the main part of the distribution  
can be represented reasonably well by a simple expression: 
\begin{equation} 
 \overline{W}(n)\,\approx\,\frac{2/\pi l}{\cosh{(2n/l)}} 
\label{eqch} 
\end{equation} 
where the parameter $l$ is related to the localization length by 
$l\,=4\pi^{-2}d$. 
If, instead of averaging the EFs with respect to their centers, we average  
them with respect to the center of the energy shell, a nice SC  
(with some tails) reappears (Fig.1b, $\beta=0.99$) in spite of  
localization . 
This shows that, in the average, the EFs homogeneously fill up 
the whole energy shell; in other words, their centers are randomly scattered 
whithin the shell (see also fig.3).   
The latter type of averaging provides a new method for calculating ergodic $d^{(e)}$, 
and hence the important localization parameters $\beta_d$ and $\lambda$ 
(\ref{eqdde}). 
  
Now we turn to the analysis, in the case $\lambda\ll 1$, of the 
other type of distribution: the GFS, or
 local spectral  
density of states $w_n(m)$, 
which is obtained from  
the columns of the matrix $C_{mn}$. The structure  
of this distribution is quite different 
from that of EFs (represented by matrix rows). Averaging  
with respect to their centers or with respect to the shell center now yields  
similar results, which  well fit the SC distribution in both cases  
(Fig.2: $\beta=0.97$ and $0.99$,  
respectively, cf. Fig.1b with $\beta=0.24$ and $0.99$). So, GFS look 
extended, yet they are localized! This is clear from the 
average of the corresponding individual $\beta$--values: $<\beta>\,=0.20$. 
The explanation of this apparent paradox is that, though each GFS is extended 
 over  
the shell,  
 it is sparse, that is, it contains many 'holes'.  
 
The difference in the structure between EFs and GFS 
is clear from Fig.3, where solid vertical bars show the main parts of EFs. 
GFS are represented by horizontal dashed lines whose sparsity immediately 
follows from scattered localized EFs.  
Our physical interpretation of the above described structure is the 
following. Spectral sparsity decreases the level density of the operative 
EFs (that is, the  ones which are actually excited in a given initial  
state). 
This is the essential mechanism of quantum localization, via decreasing 
the relaxation time scale \cite{CaCh}\cite{CaChCo}. Yet, the initial diffusion and  
relaxation are still classical, similar to the ergodic case, which requires 
extended GFS. On the other hand, EFs are directly 
related to the steady--state distribution, both being solid because of the  
homogeneous diffusion during the statistical relaxation. 

In conclusion, we have analyzed the structure of the GFS 
(local spectral density  
of states) and of the eigenfunctions for a class of Random Matrices  
which comes much closer to the structure of the real Hamiltonian matrix  
of a conservative system than the conventional full Random Matrices. We have provided  
numerical evidence for the existence of both a delocalized regime, in  
which eigenfunctions have maximal size, with an average shape close to  
the semicircle law, and of a localized regime, in which the size of EFs is  
much smaller than the semicircle width. 
More precisely, quantum localization introduces a symmetry breaking, in the 
sense that the eigenfunctions are solid, narrow and randomly scattered inside 
the energy shell, while the GFS remain extended over the whole shell but 
become sparse. In classical language, the  
latter situation means that, although classical trajectories are  
ergodically distributed over the whole energy surface, the quantum  
eigenfunctions cover but a small region of the latter. 
Thus our results indicate that quantum localization is a more general  
phenomenon  
than commonly believed, and suggest similar investigations for  
realistic Hamiltonian, conservative, classically chaotic systems. 
 
Support by the NATO linkage grant LG930333 and by I.N.F.M. is acknowledged. 
Partial 
support by grant RB7000 is also gratefully acknowledged by F.M.I.
 
 B.V.C. and 
F.M.I. are grateful to their colleagues of the University of Milano at Como
for their hospitality during the period in which this work was completed. 

\begin{figure}\caption{Structure of ergodic (1a) and localized (1b)  
eigenfunctions. Each figure corresponds to a single matrix 
with parameters $N=2560$,$v=0.1$, 
$b=16$,$\rho=40$,$q=(\rho v)^2/b=1$ (a), and $N=2400$,$v=0.1$,$b=10$, 
$\rho=300$, $q=90$ (b). The fat full line is the semicircle law (5).  
Solid lines were obtained by averaging $300$ eigenfunctions with respect to  
their centers; circles, by averaging the same eigenfunctions with respect to  
the centers of their energy shells. In the ergodic case (a) $\lambda=3.7$,  
all distributions are close to one another apart from fluctuations. In the  
localized case (b) $\lambda=0.24$, the average with respect to centers 
$n_c(m)$ of the distributions $W_m(n)$ shows  
a clear localization with $\beta=0.24$, 
while the other average remains close  
to semicircle, with $\beta=0.99$.} 
\end{figure} 
 
\begin{figure}\caption{Structure of the GFS (local DOS) for a single  
matrix, with the same parameters as for fig.1b. The same averages as in Fig.1 
are shown, and unlike that case they are close to each other and to the  
semicircle law.} 
\end{figure} 
 
\begin{figure}\caption{A comparison of the structure of eigenfunctions and of  
GFS in the localized case of fig.1b. Solid vertical bars represent the  
widths $\Delta n$ of individual eigenfunctions over the unperturbed basis.   
Horizontal dotted lines show the size $\Delta m$ of the local spectrum for  
individual basis states. 
 Although all basis states have comparable sizes,  
close to the size of the energy shell, they are very sparse ($\beta=0.20$),  
due to the fact that   
EFs are strongly localized, and irregularly scattered inside the energy shell. 
} 
\end{figure}


\begin{thebibliography}{99} 
\bibitem{bible}G.Casati, B.V.Chirikov, J.Ford and F.M.Izrailev,
{\it Lect. Notes in Physics},{\bf 93},(1979), 334 (also in ref.6).
\bibitem{W} E.Wigner, Ann. Math. {\bf 62}, 548 (1955); {\bf 65}, 203 (1957). 
\bibitem{BM} T. Brody, J. Flores, J. French, P. Mello, A. Pandey, and 
      S. Wong, Rev. Mod. Phys. {\bf 53}, 385 (1981); 
      M. Mehta, {\it Random Matrices }, (Academic Press, New York,1991). 
\bibitem{Ch} B.V. Chirikov, Phys. Lett. A {\bf 108}, 68 (1985). 
\bibitem{FG} V.V. Flambaum, A.A. Gribakina, G.F. Gribakin, and M.G. Kozlov, 
      Phys. Rev. A {\bf 50}, 267 (1994). 
\bibitem{CaCh} G. Casati and B.V. Chirikov, The legacy of chaos  
 in quantum mechanics, in: {\it Quantum Chaos: Between Order 
        and Disorder}, edited by G. Casati and B.V. Chirikov 
               (Cambridge University Press, Cambridge, 1995). 
\bibitem{Flam} V.V.Flambaum, F.M.Izrailev and G.Casati, preprint DYSCO58  
               Como 1995. 
 
\bibitem{CCGI} G. Casati, B.V. Chirikov, I. Guarneri and F.M. Izrailev, 
               Phys. Rev. E {\bf 48}, R1613 (1993). 
\bibitem{CaChCo} G. Casati and B.V. Chirikov, Physica D, {\bf 86} (1995) 220.

\bibitem{YaF} Y.V. Fyodorov, O.A.Chubykalo, F.M.Izrailev, G.Casati, 
               "Wigner random banded matrices with sparse structure: 
              Local spectral density of states", preprint DYSCO 95. 

\bibitem{Fe} D.M.Leitner and M.Feingold, J.Phys. A {\bf 26} (1993) 7367.
\bibitem{FLW} the parameter $\lambda$ was introduced to describe energy level 
 statistics in: 
M. Feingold, D. Leitner, M. Wilkinson, Phys. Rev. Lett. 
              {\bf 66}, 986 (1991); J. Phys. A {\bf 24}, 1751 (1991). 
 
\end{thebibliography}
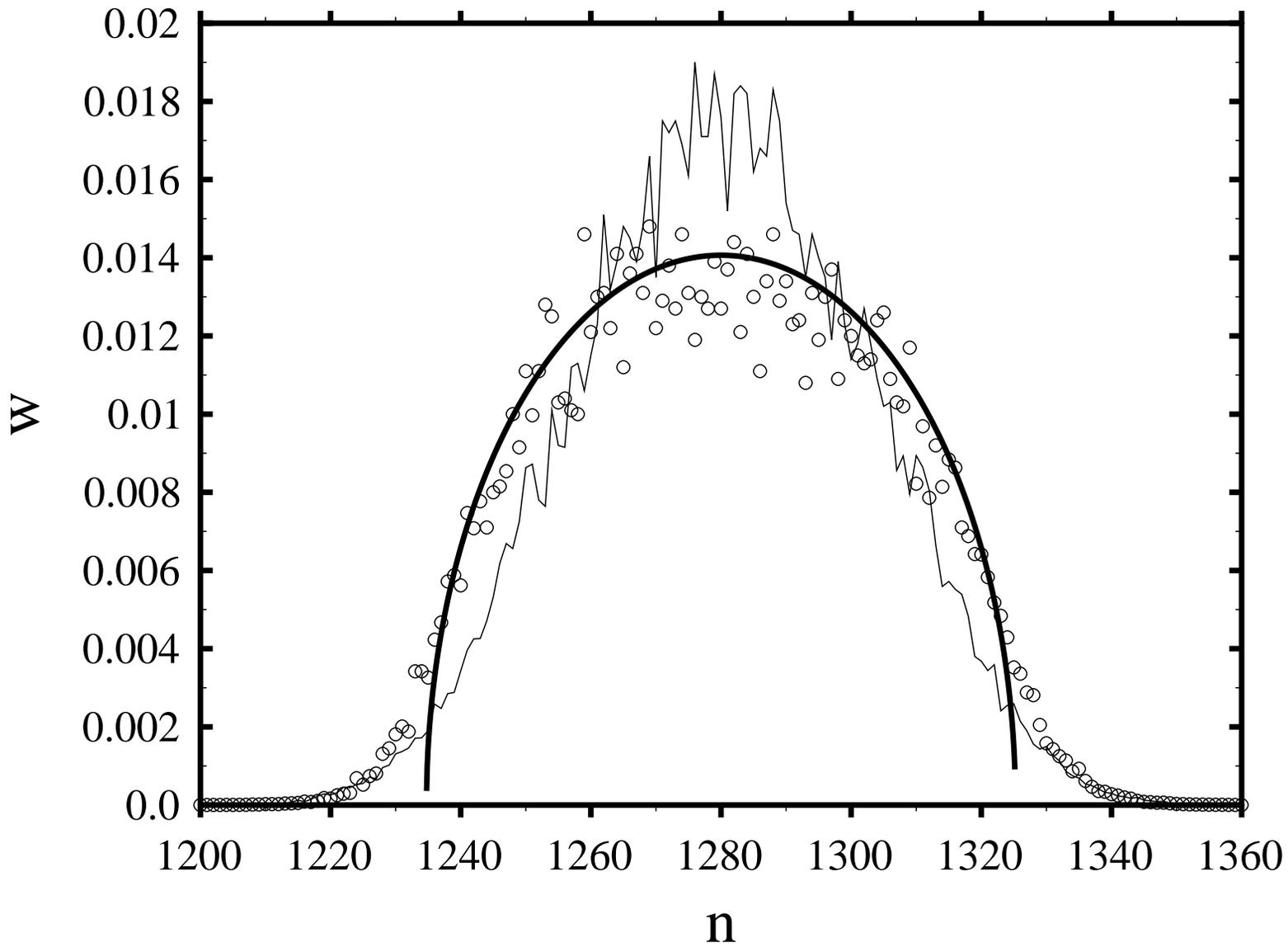
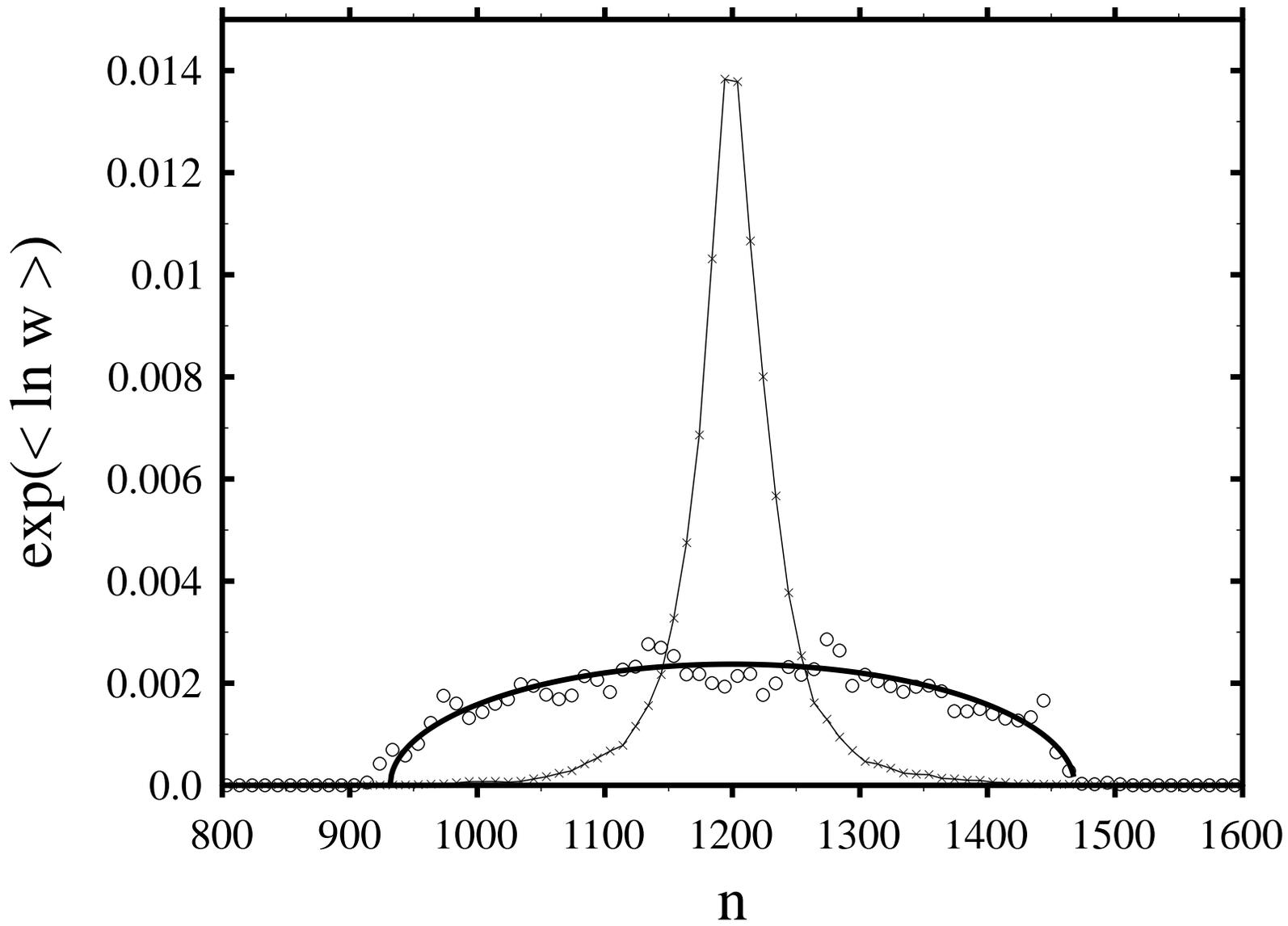
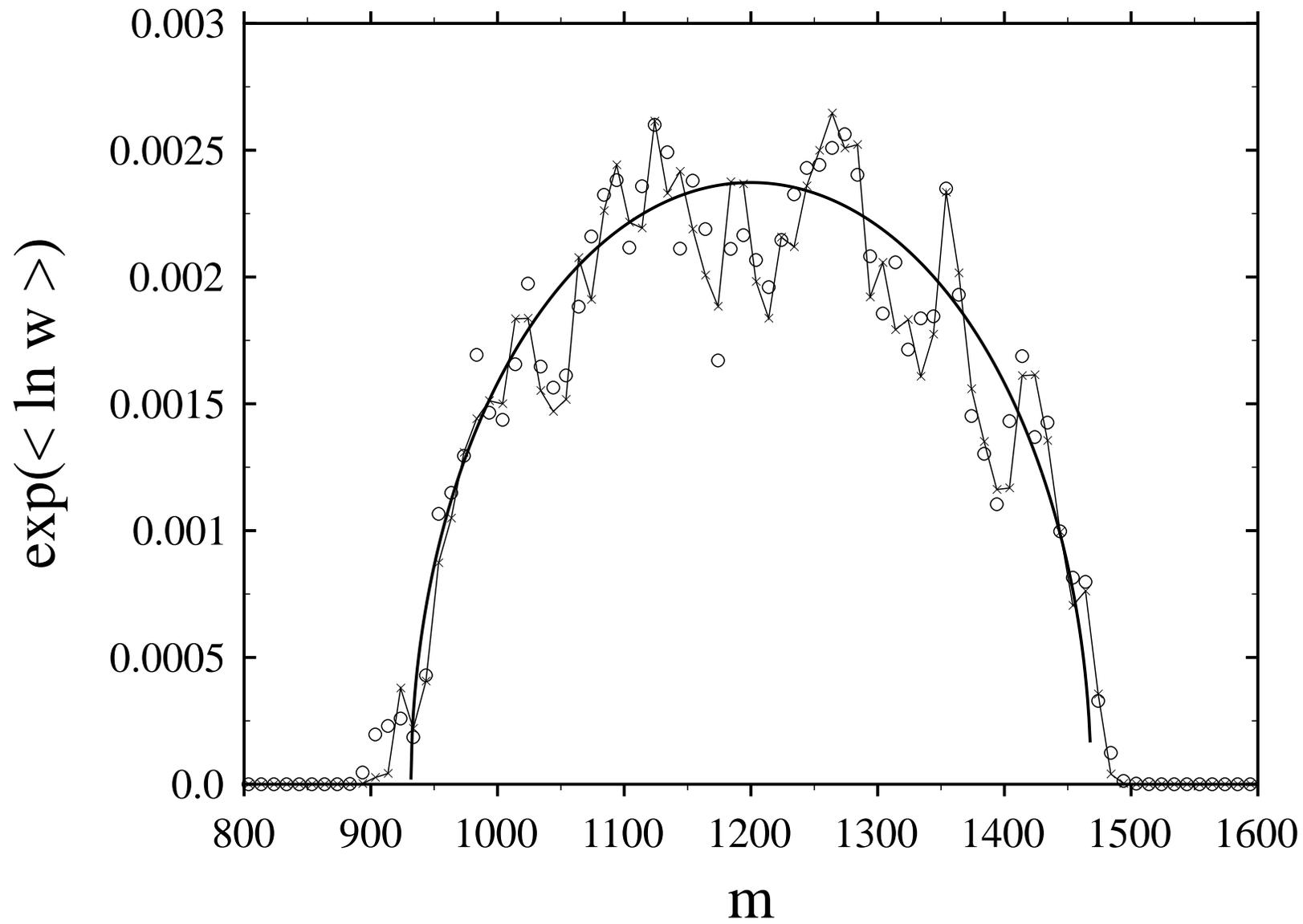
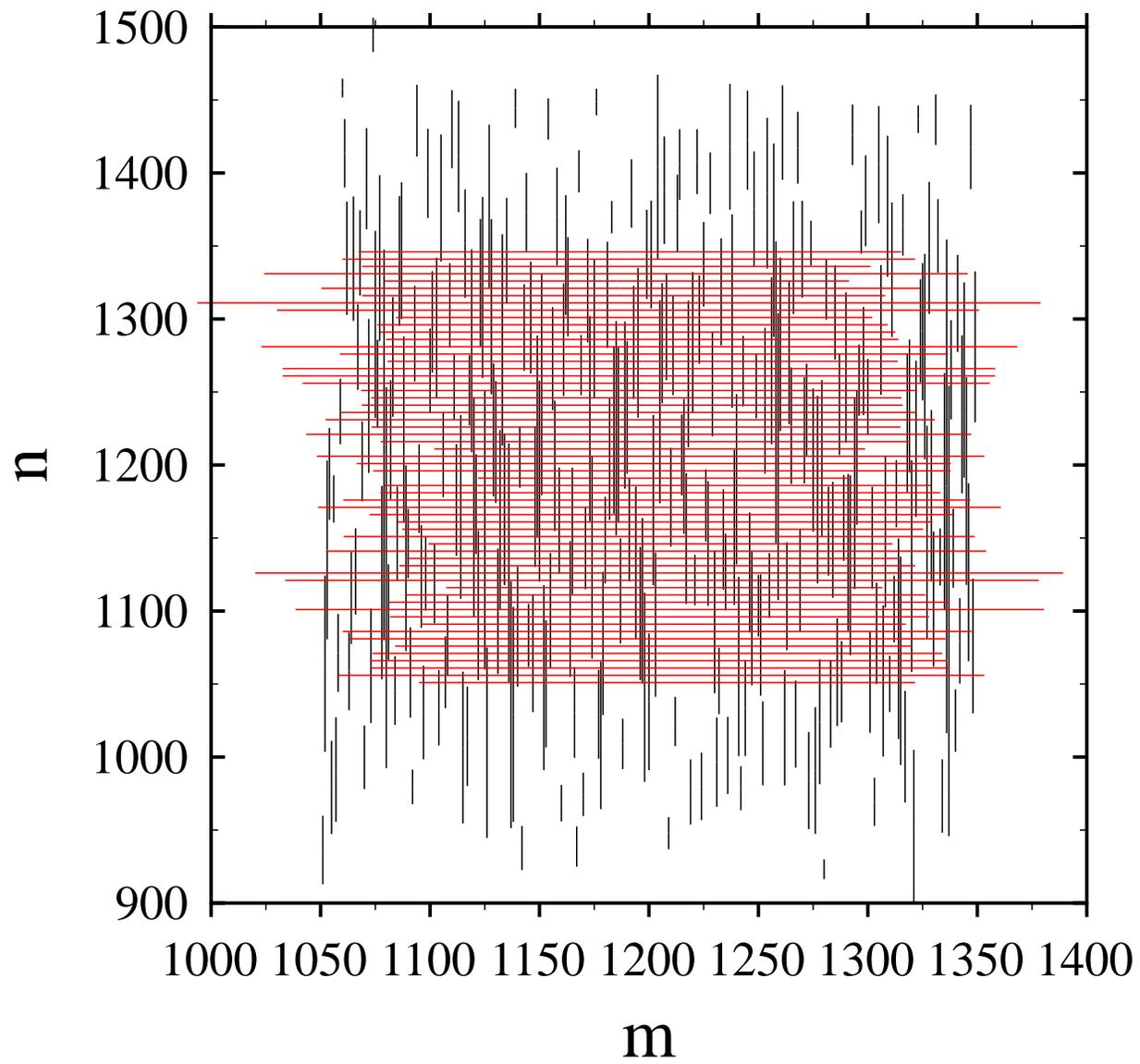
\end{document}